\documentclass[a4paper,10pt]{article}
\usepackage{fullpage}
\usepackage{authblk}
\usepackage[numbers,sort&compress]{natbib}

\usepackage[utf8]{inputenc}
\usepackage{amsmath}
\usepackage{amsfonts}
\usepackage{amssymb}
\usepackage[pdftex]{hyperref}
\usepackage{graphicx}
\usepackage{subfigure}
\usepackage{algorithmic}
\usepackage{algorithm}
\usepackage{color}
\usepackage{bm}

\title{Multifractal cross-correlation effects in two-variable time series of complex network vertex observables}

\author[1]{Pawe\l{} O\'{s}wi\c{e}cimka\thanks{pawel.oswiecimka@ifj.edu.pl}}
\author[2]{Lorenzo Livi\thanks{llivi@scs.ryerson.ca}\thanks{Corresponding author}}
\author[1,3]{Stanis\l{}aw Dro\.{z}d\.{z}\thanks{stanislaw.drozdz@ifj.edu.pl}}
\affil[1]{Complex Systems Theory Department, Institute of Nuclear Physics, Polish Academy of Sciences, PL-31-342 Krak\'{o}w, Poland}
\affil[2]{Dept. of Computer Science, College of Engineering, Mathematics and Physical Sciences, University of Exeter, Exeter EX4 4QF, UK}
\affil[3]{Institute of Computer Science, Faculty of Physics, Mathematics and Computer Science, Cracow University of Technology, PL-31-155 Krak\'{o}w, Poland}
\providecommand{\keywords}[1]{\textbf{\textit{Keywords---}} #1}

\begin{document}

\maketitle

\begin{abstract}
We investigate the scaling of the cross-correlations calculated for two-variable time series containing vertex properties in the context of complex networks. Time series of such observables are obtained by means of stationary, unbiased random walks.
We consider three vertex properties that provide, respectively, short, medium, and long-range information regarding the topological role of vertices in a given network.
In order to reveal the relation between these quantities, we applied the multifractal cross-correlation analysis technique, which provides information about the nonlinear effects in coupling of time series. We show that the considered network models are characterized by unique multifractal properties of the cross-correlation. In particular, it is possible to distinguish between Erd\"{o}s-R\'{e}nyi, Barab\'{a}si-Albert, and Watts-Strogatz networks on the basis of fractal cross-correlation.
Moreover, the analysis of protein contact networks reveals characteristics shared with both scale-free and small-world models.\\
\end{abstract}

\section{Introduction}

Complex biological, social, and technological systems are everywhere around us, from cellular environments to natural language and financial markets \cite{tkavcik2014information,bialek2015perspectives,kwapien2012physical,havlin1999scaling}.
The field of network science \cite{newman2003structure,albert2002statistical,dorogovtsev2008critical} offers a well-established mathematical setting to study complex systems made of units (vertices) that interact in some way (binary relations; edges).
In fact, networks can be characterized in a multitude of ways, using either structural and dynamical features \cite{boccaletti+latora+moreno+chavez+hwang2006}.
Assortativity in networks \cite{newman2002assortative} refers to the tendency of adjacent vertices to share a similar property, such as assortativity or disassortativity of vertex degree. Assortativity has been used for characterizing both models and real networks and it is known to convey useful information describing the organization of a network \cite{newman2003mixing}.

Recent research endeavors have highlighted a possible interplay between time series and complex networks \cite{campanharo2011duality}.
In particular, it is suggested that time series can be mapped to a network and then analyzed by using methods devised for complex networks \cite{donner2011,lacasa2008time,subramaniyam2015signatures}.
The dual approach consists in mapping a complex network to a time series and performing the analysis accordingly \cite{nicosia2013characteristic,mixbionets2,weng2014time}.
In the latter case, the mapping can be performed by means of the framework of random walks on graphs \cite{rosvall2011multilevel,burioni2005random,gallos2007scaling}. At each time step, a vertex property is ``emitted'' by the walker, resulting in a time series that encodes the structural organization of the network from the point of view of the particular vertex property under analysis.

In \cite{weng2014time} the authors suggest the existence (although without proof) of a monotonically increasing relationship between assortativity and the degree of persinstency computed on a time series representation of the vertex property under analysis.
Recently, H{\"o}ll and Kantz \cite{holl2015relationship} demonstrated the equivalency between the autocorrelation and the detrended fluctuation analysis, which is a well-known procedure to assess the degree of persistency of time series with trends (for previous related works on the same topic, see \cite{kantelhardt2001detecting,heneghan2000establishing,jaffard1997multifractal}).
Therefore, a reasonable inference step would consist in linking the assortativity degree of a vertex property with the related time series autocorrelation.
Such a hypothesis is supported by experimental results on several different networks representing biological and technological systems \cite{nicosia2013characteristic,mixbionets2}.

In this paper, we elaborate over these ideas and extend the analysis framework to cross-correlations \cite{oswiecimka2014detrended}.
Notably, we study the scaling and related multifractal properties of cross-correlations of time series related to two different vertex properties (e.g., cross-correlations between degree and clustering coefficient) of the same network.
As mentioned before, persistency of time series finds an analogue interpretation in terms of assortativity in networks.
However, to the best of our knowledge, there is no measure for complex networks that captures the notion of cross-correlation scaling between two different vertex properties.
This stresses the novelty of our approach with respect to previous single-variable studies \cite{nicosia2013characteristic,mixbionets2,weng2014time}.
We discuss the results of the simulations performed on both network models and real networks representing folded proteins \cite{ecoli_graph}.
Our results suggest that the proposed analysis framework is useful to investigate the multi-level organization of networks from the perspective of coupling between two different topological features of vertices.
In particular, differently from what emerged in single-variable studies, we show that analyzing two-variable time series allows to distinguish even between network models with uncorrelated vertex properties (e.g., between scale-free and Erd\"{o}s-R\'{e}nyi graphs). In addition, analysis of cross-correlations highlights the presence of structural constraints existing in protein contact networks and suggests an underlying structural dualism possessing both scale-free and small-world properties.

\section{Single and two-variable scaling of fluctuations}
\label{sec:fluctuations}

Self-similarity of time series can be investigated by focusing on the scaling properties of the fluctuations.
Several different methods have been proposed to analyze the fluctuations of the stationary component of a time series \cite{riley2012tutorial,fernandez2013measuring,wendt2007multifractality}.
The Detrended Fluctuation Analysis (DFA) procedure is a state-of-the-art method conceived for this purpose.
Its generalization, called Multifractal Detrended Fluctuation Analysis (MF-DFA) \cite{kantelhardt2002multifractal,PhysRevE.74.016103}, accounts also for multi-scaling, allowing a multi-level characterization of time series.

However, the analysis of each time series separately could be insufficient to comprehensively characterize the system and to capture the subtleties involved with its temporal (or space) organization. From this point of view, nonlinear (multifractal) cross-correlation between time series representing different quantities turns out to be crucial 
\cite{oswiecimka2014detrended,kwapien2015detrended,rak2015detrended,zhou2008multifractal,kristoufek2015can}.
In this study, we use an algorithm called Multifractal Cross-Correlation Analysis (MF-CCA), whose functioning can be described as follows.
Given two time series $x$ and $y$, it is possible to formulate a function $F^{q}_{XY}(\cdot)$: 
\begin{equation}
\label{eq:fluctuations_crosscorr}
F^{q}_{XY}(s) = \frac{1}{2N_s} \sum_{\nu=1}^{2N_s} \mathrm{sign}(F^{2}_{XY}(s, \nu))|F^{2}_{XY}(s, \nu)|^{q/2},
\end{equation}
where $N_s$ is the number of disjoint segments $\nu$ of length $s$, $X$ and $Y$ denote, respectively, the detrended version of $x$ and $y$ (here, we removed a fitted second-degree polynomial), $\mathrm{sign}(\cdot)$ is the sign function, and $F^{2}_{XY}(s, \nu)=1/s\sum_{i=1}^{s}X_{\nu}(s, i)Y_{\nu}(s, i)$.
In Eq. \ref{eq:fluctuations_crosscorr}, $q$ acts as a lens magnifying the contribution of small/large fluctuations. In particular, when $q<0$ the focus is on small fluctuations, while for $q>0$ on the large ones.
When only one time series is taken into account, $F_{XX(YY)}(q,s)$, we retrieve MF-DFA fluctuation function.
Power-law scaling of the fluctuation function and covariance is manifested by the following scaling relation: $[F^{q}_{XY}(s)]^{1/q}=F_{XY}(q,s)\sim s^{\lambda_q}$ and $F_{XX(YY)}(q,s)\sim s^{h_{x(y)}(q)}$, correspondingly, where $h_{x(y)}(q)$ stands for generalized Hurst exponent of either $x$ or $y$.
If $F^{q}_{XY}(s)$ does not develop a power-law scaling (e.g., fluctuating around zero), then there is no fractal cross-correlation between the time series under study for the particular $q$ value.
In the case of negative cross-correlations (if $q$th-order covariance function is negative for every $s$), it is possible to consider $F^{q}_{XY}(s)=-F^{q}_{XY}(s)$.
Similarly to the generalized Hurst exponent, $\lambda_q$ is expected to be independent of $q$ in the case of monofractal cross-correlations; multifractal cross-correlations are instead highlighted by the $q$-dependency of $\lambda_q$.
Relation between $\lambda_q$ and bivariate Hurst exponent, $h_{xy}(q)=[h_x(q)+h_y(q)]/2$, provides additional information on cross-correlations behavior. 
In particular, the larger is the difference between $\lambda_q$ and $h_{xy}(q)$, the more mutually uncorrelated is the (multi-)fractal organization of the two series \cite{oswiecimka2014detrended}.

It also needs to be stressed that, in general, the MF-CCA procedure involves probing not only cross-correlations between the corresponding series of the pointwise H\"{o}lder exponents but it also accounts for the relative signs of fluctuations and even for the relative ratios of their amplitudes. MF-CCA thus sets maximally stringent requirements on identification of the (multi-)fractal coherence among the time series. This has been explicitly verified on several models as well as on some realistic cases \cite{oswiecimka2014detrended} and also tested, for the purpose of the present study, on some further, multifractional Brownian motion time series generated from the pointwise H\"{o}lder exponents \cite{levy1995multifractional,ayache2004identification,trujillo2012evolving}.

\section{Random walks on graphs and related time series of vertex properties}
\label{sec:rw}

Let $G=(\mathcal{V}, \mathcal{E})$ be an undirected graph, where $\mathcal{V}$ and $\mathcal{E}$ denote vertices and edges, respectively.
A Markovian random walk in a graph \cite{Dehmer201157} is a first-order Markov chain that generates sequences of vertices.
Transitions among vertices/states are regulated by the transition matrix,
\begin{equation}
\label{eq:transition_matrix}
\mathbf{T}=\mathbf{D}^{-1}\mathbf{A},
\end{equation}
where $\mathbf{A}$ is the adjacency matrix of $G$ and $\mathbf{D}$ is a diagonal matrix of vertex degrees, i.e., $D_{ii}=\mathrm{deg}(v_i)=\sum_{j} A_{ij}$.
Matrices in (\ref{eq:transition_matrix}) are intended to have a size equal to the cardinality of $\mathcal{V}$.

Let $\mathbf{p}_{0}$ be the initial distribution of the chain.
The probability of the states at time $t>0$ can be computed as $\mathbf{p}_{t}=\mathbf{p}_{t-1}\mathbf{T}=\mathbf{p}_{0}\mathbf{T}^{t}$.
The stationary distribution, $\boldsymbol\pi$, is a probability vector satisfying $\boldsymbol\pi=\boldsymbol\pi\mathbf{T}$, i.e., it can be intended as the limiting distribution of the vertices/states when $t\rightarrow\infty$.
If the graph is undirected and non-bipartite, then a random walk always possesses a (unique) stationary distribution: $\pi_i = \mathrm{deg}(v_i)/(2|\mathcal{E}|)$.

Let $M_{P}: \mathcal{V}\rightarrow\mathcal{O}_{P}$ be a time-homogeneous vertex property map, where $\mathcal{O}_{P}$ is the domain of vertex property $P$, such as degree, clustering coefficient or other user-defined vertex properties.
By performing a (stationary) random walk on $G$, a sequence of vertices $(v_{i_{0}}, v_{i_{1}}, ..., v_{i_{T}})$ is produced.
We can associate to such a sequence another sequence of observables by emitting at each time instant $t$ the corresponding vertex properties given by $O_t=M_{P}(v_{i_{t}})$, where $v_{i_{t}}$ is the vertex visited at time $t$.
This process generates a sequence of vertex properties $\mathcal{S}_{P}=(O_0, O_{1}, ..., O_{T})$, which can be seen as a time series, since the values are equally spaced in time.
A first-order Markov process is by definition a memory-less process. Nonetheless, the time series $\mathcal{S}_{P}$ might posses a complex organization characterized by a non-trivial correlation structure.

A previous study \cite{nicosia2013characteristic} revealed the presence of a cross-over for the scaling function in such time series $\mathcal{S}_{P}$.
The authors noted that the scales affected by the cross-over depend on both the vertex property and type of graph.
However, to the best of our knowledge, there is no analytical result explaining this fact.
Nonetheless, we conjecture that the scales affected by the cross-over are mostly determined by the strength of the assortativity of the vertex property $P$ under consideration.

Single-variable studies aimed at analyzing the (multi-)fractal properties of such time series $\mathcal{S}_{P}$ show the possibility to observe non-trivial organizations in both technological and biological networks \cite{mixbionets2,nicosia2013characteristic,weng2014time}.
Here, we build on these studies and analyze the scaling properties of the cross-correlations between two time series, say $\mathcal{S}_{P_{1}}$ and $\mathcal{S}_{P_{2}}$, composed of two distinct vertex properties $P_{1}$ and $P_{2}$ of the same graph $G$.

\section{Simulations}

In all experiments, we consider three vertex properties for the time series: degree (VD), clustering coefficient (VCL), and closeness centrality (VCC).
Such properties describe short, medium, and long-range topological information of a vertex in the corresponding network, respectively.
The length of the random walks is defined as $10^6$ time steps, which takes into account the number of vertices in the networks taken into account.
The analysis is performed by means of the MF-CCA procedure \cite{oswiecimka2014detrended}.
For each network, we instantiate 40 different (independent) random walks. Results of MF-CCA calculations are accordingly reported as averages with related standard deviations.

\subsection{Synthetic network models}
\label{sec:synth_models}

\begin{figure}[htp!]
\centering

\begin{tabular}{cc}
\includegraphics[scale=0.27,keepaspectratio=true]{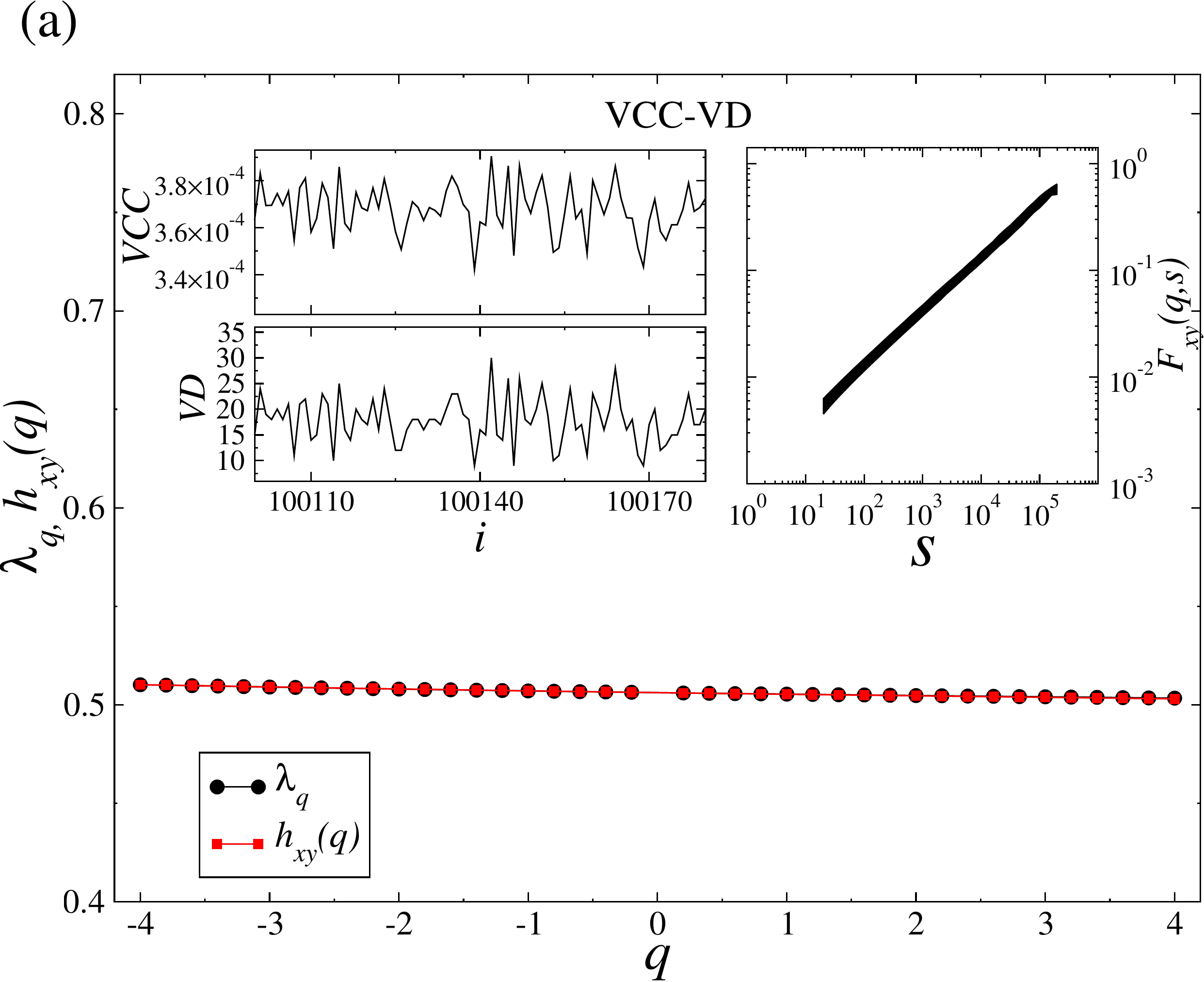}
&
\includegraphics[scale=0.27,keepaspectratio=true]{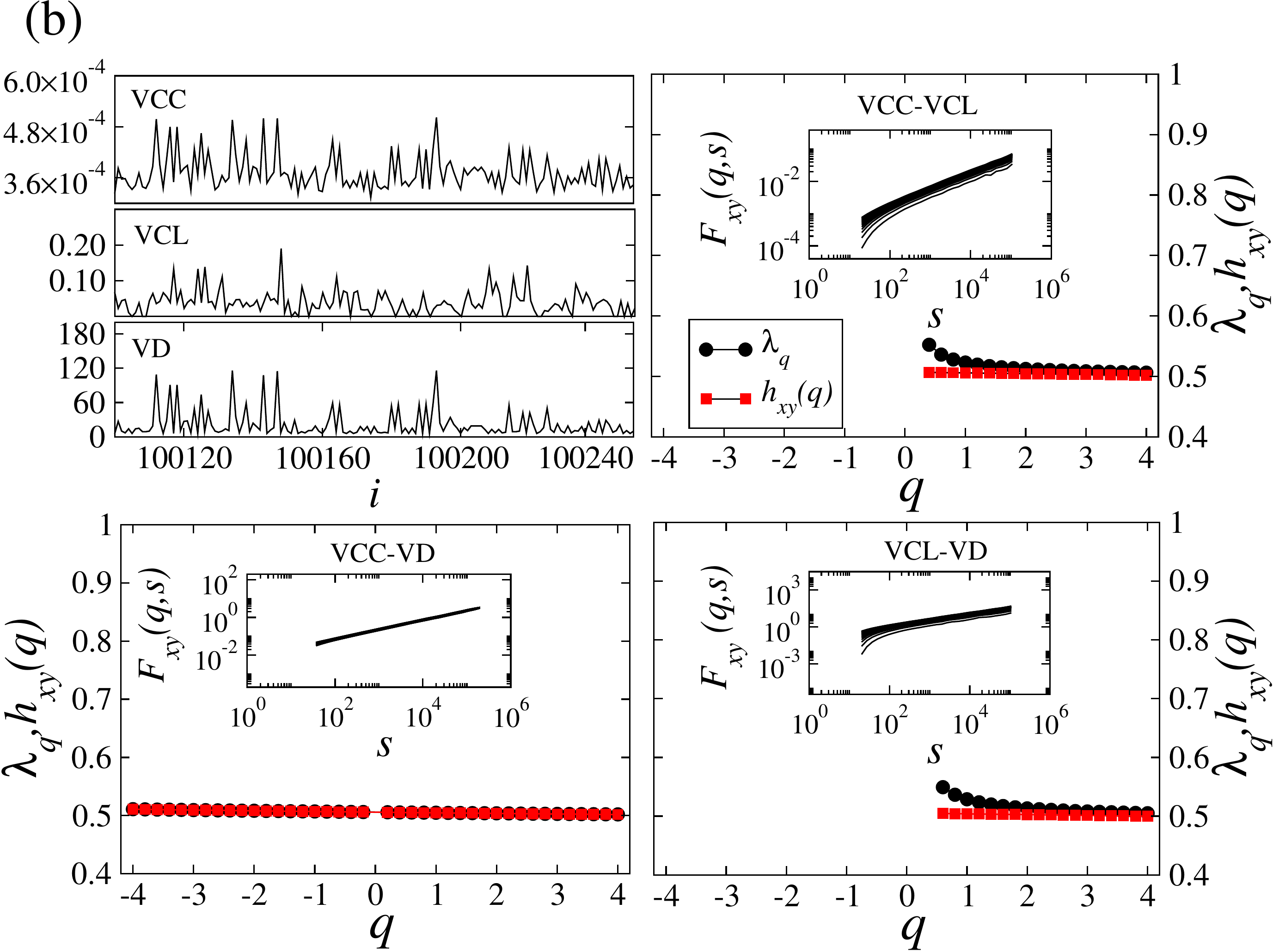}
\end{tabular}

\caption{Results for ER (a) and BA (b) networks. Results for VCC-VD are comparable. In the BA case, cross-correlations can be computed also for VCC-VCL and VCL-VD for positive $q$.}
\label{fig:ER_BA}
\end{figure}

Synthetic graphs are always generated with 1000 vertices.
First, we take into account Erd\"{o}s-R\'{e}nyi (ER) and Barab\'{a}si-Albert (BA) graphs.
ER graphs are generated with edge probability equal to 0.017; BA graphs by attaching each new vertex with 6 old vertices.
Fig. \ref{fig:ER_BA} shows the results.
For ER graphs, it was possible to obtain power-law scaling of cross-correlations only in the VCC-VD case (Fig. \ref{fig:ER_BA} panel (a)).
For the other two cases, i.e., VCC(VD)-VCL pairs, $q$th-order covariance functions $F^{q}_{XY}(s)$ determined by Eq. \ref{eq:fluctuations_crosscorr} fluctuates strongly around zero. This means that clustering coefficient is not fractally correlated with vertices degree and centrality.

In fact, the relation between the two quantities can be appreciated even by visual inspection of inset of Fig. \ref{fig:ER_BA} panel (a) showing the corresponding time series.
The observed agreement between $\lambda_q$ and bivariate Hurst exponent $h_{xy}(q)$ indicates strong connection between VD related VCC values: the higher the vertex degree, the smaller the topological distances (larger centrality).
Monofractality of the cross-correlation characteristic $\lambda_q$ (black, filled circles) is the result of monofractal organization of the corresponding single-variable time series ($h_{x(y)}(q) \approx  0.5$). In fact, in ER networks assortativity of each considered quantity is close to zero, which is confirmed by the value of Hurst exponent $h_{x(y)}(2) \approx  0.5$ estimated for the corresponding single-variable time series (not shown).

\begin{figure}[htp!]
\centering

\begin{tabular}{cc}
\includegraphics[scale=0.175,keepaspectratio=true]{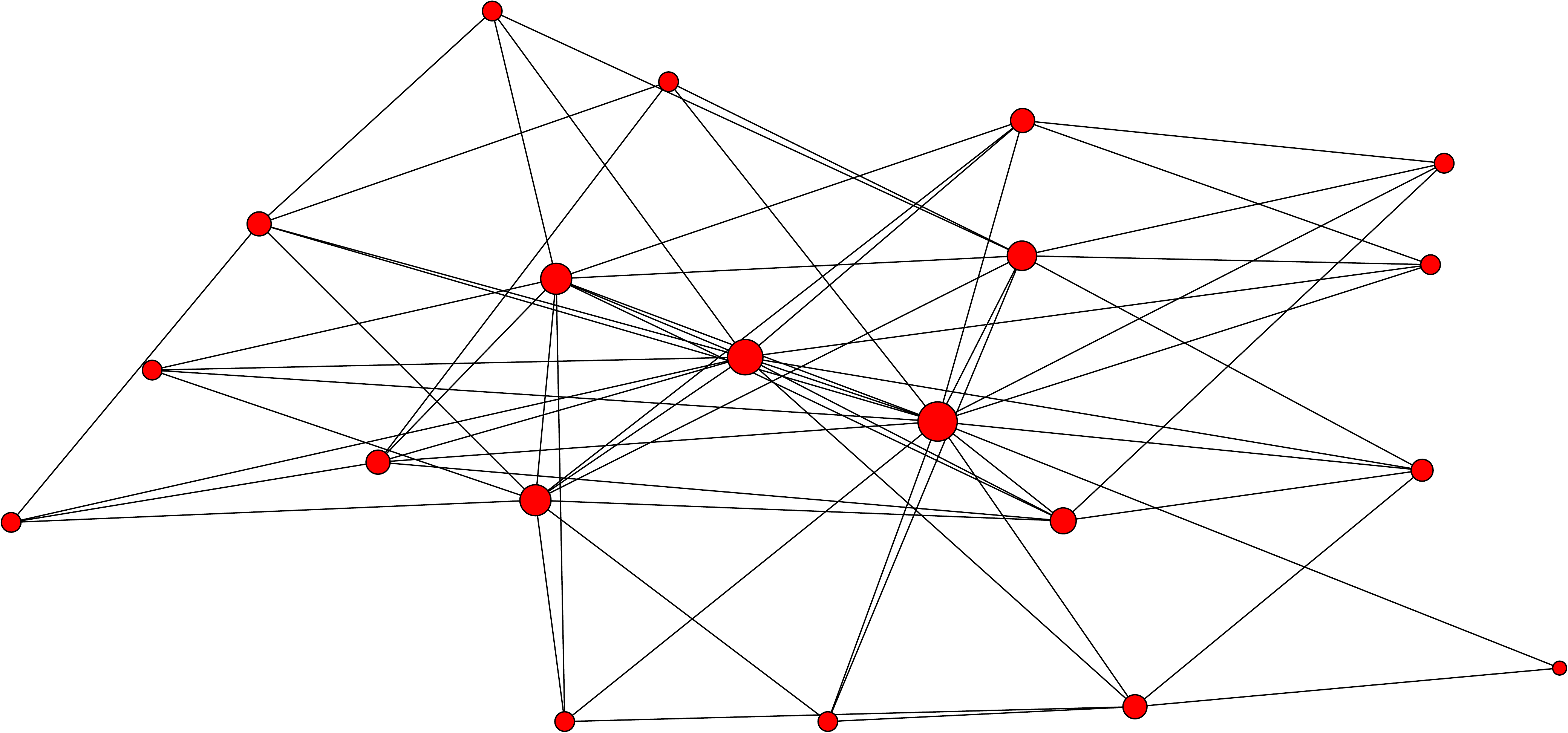}
&
\includegraphics[scale=0.175,keepaspectratio=true]{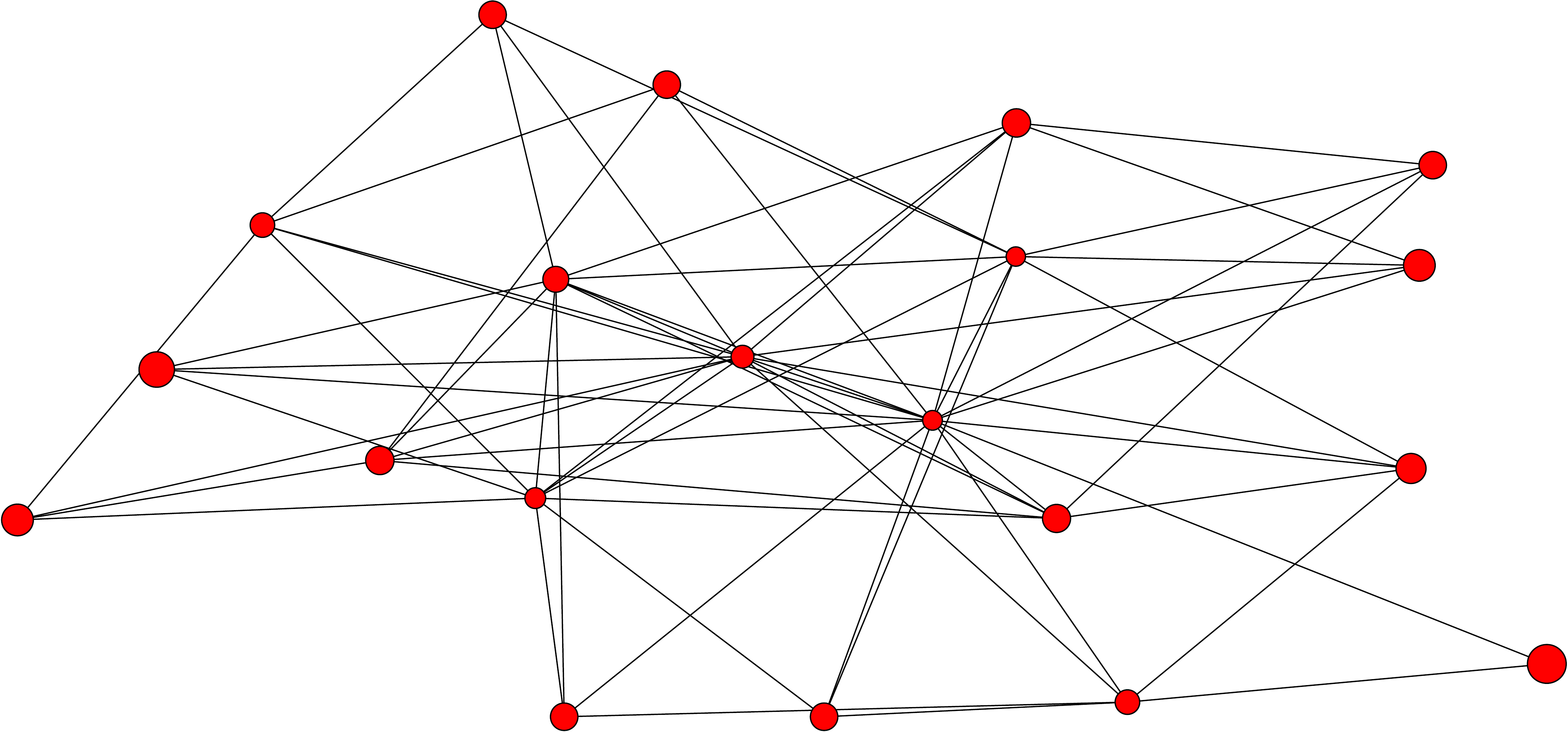}
\end{tabular}

\caption{Graphical representation of a small BA network with 20 vertices. In the left-hand panel, size of vertices is proportional to vertex degree. In the right-hand panel, instead, size of vertices is proportional to vertex clustering coefficient. As it is possible to notice, hubs are characterized by small clustering coefficient. On the other hand, hubs are directly connected with vertices having large clustering coefficient. This does not happen in ER networks, since they are characterized by very low average clustering coefficient.}
\label{fig:BAdetail}
\end{figure}

In BA network, fractal properties of VCC, VCL, and VD time series are characterized by a single scaling exponent $h_{x(y)}(q) \approx  0.5$. Therefore, from the classical DFA technique point of view, the distinction between ER and BA models is impossible. However, the difference between such graphs is revealed on the level of multifractal cross-correlation analysis.
Relation between VCC-VD times series is similar to that obtained for ER network, although, in this case, it can be explained with more clarity.
In fact, BA networks contain hubs characterized high vertex degree and (closeness) centrality.
However, in the BA case, it is possible to obtain results also for cross-correlations of VCC-VCL and VCL-VD (Fig. \ref{fig:ER_BA} panel (b)). 
The standard cross-correlation function (with unitary time lag) indicates relation between clustering and vertex degree of adjacent vertices as a source of fractal cross-correlation (results not shown).
Moreover, the MF-CCA cross-correlation is well-defined only for large fluctuations (characterized by positive $q$), which are related to hubs (large VD and VCC) and many neighbouring nodes that are connected (large VCL).
Fig. \ref{fig:BAdetail} provides a graphical illustration of this result.

\begin{figure}[ht!]
\centering

\begin{tabular}{cc}
\includegraphics[scale=0.27,keepaspectratio=true]{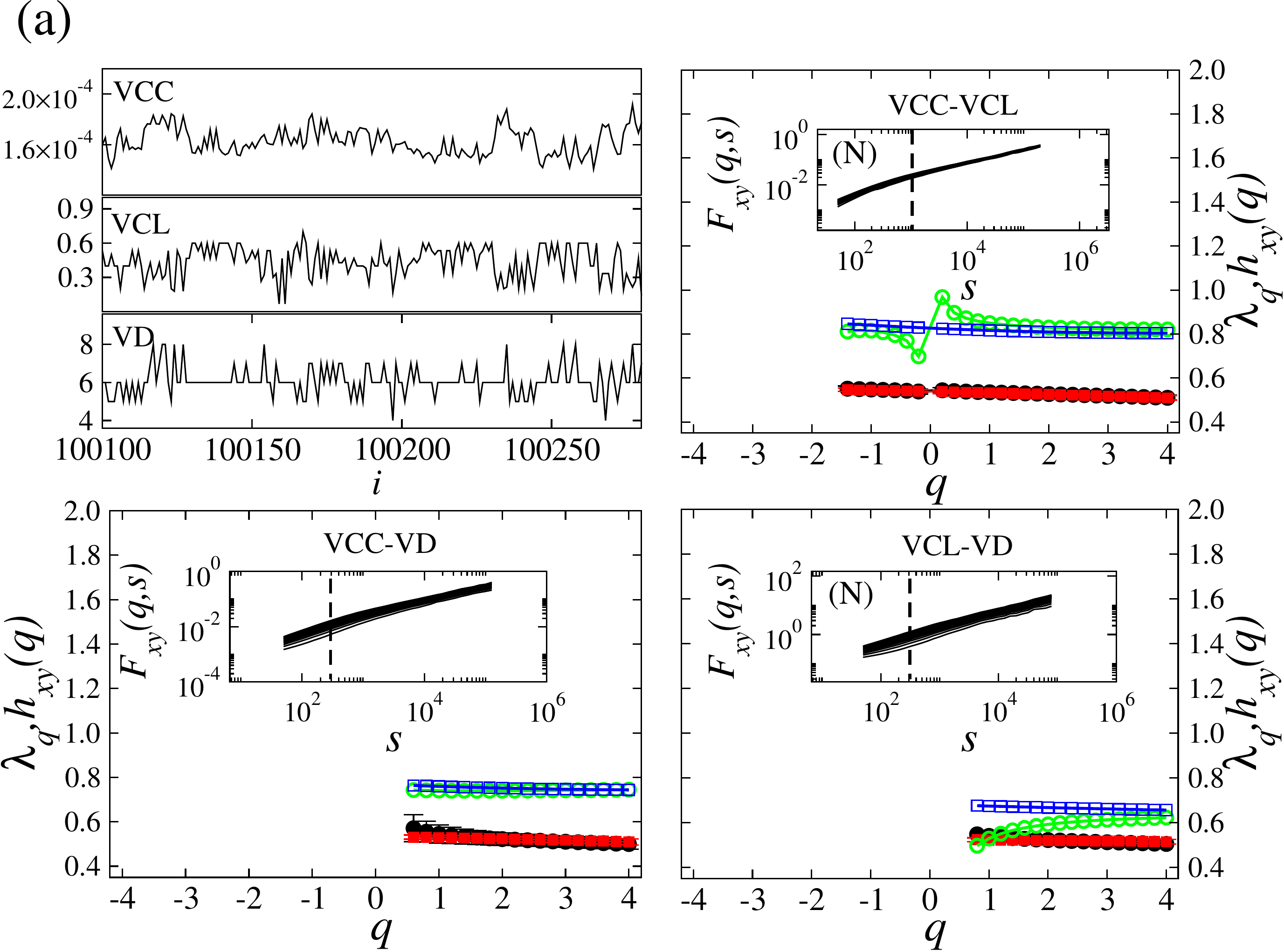}
&
\includegraphics[scale=0.27,keepaspectratio=true]{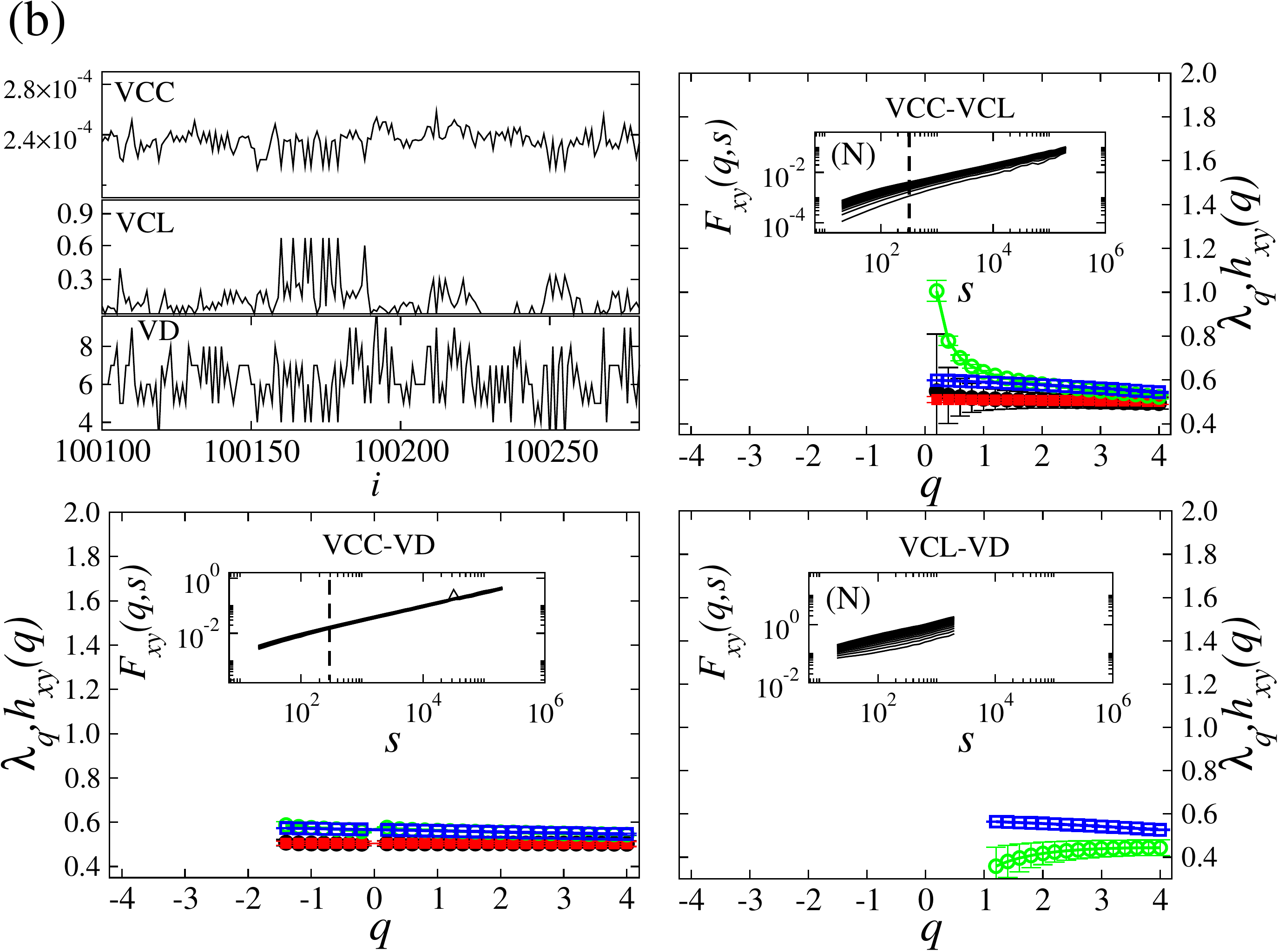}\\
\multicolumn{2}{c}{\includegraphics[scale=0.3,keepaspectratio=true]{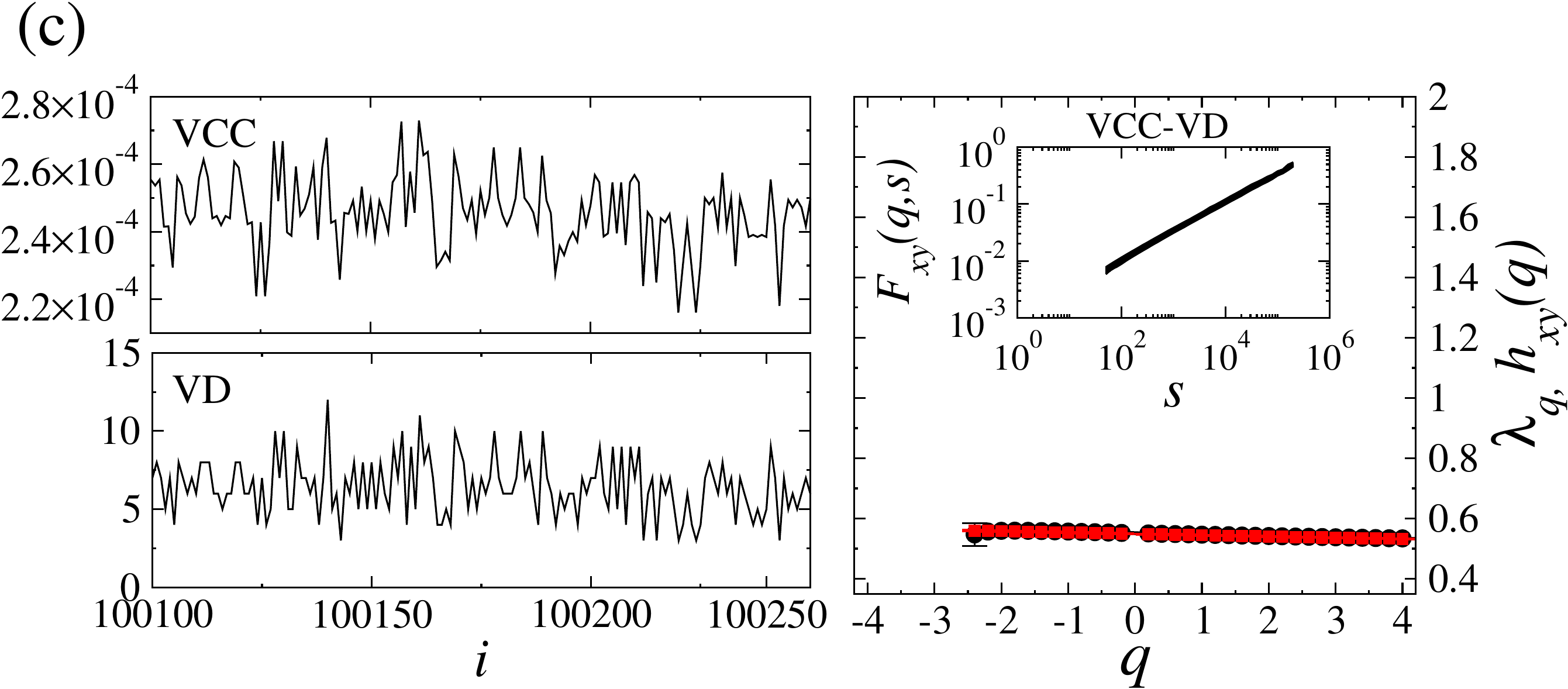}}
\end{tabular}

\caption{Results for WS networks with $k=6$ and $p=0.1$ (a), 0.5 (b), and 0.9 (c). We show time series excerpts, $F_{XY}(q, s)$, and $\lambda_q$ (circles), $h_{xy}(q)$ (squares). The empty (green and blue) and full (red and black) symbols refer to the short- and long-range correlations, respectively. The vertical dashed lines in the insets indicate the cross-over between different scaling regimes.}
\label{fig:WS}
\end{figure}

Fig. \ref{fig:WS} shows the results obtained for the Watts-Strogatz (WS) model.
Two parameters control the model: $k$, number of nearest neighbors in the initial ring-like topology, and $p$, the probability of rewiring an edge considering a uniformly chosen pair of vertices. We show results for WS models with $k=6$ and $p$ equal to 0.1, 0.5, and 0.9.

For $p=0.1$, we observe fractal cross-correlation between all considered pairs of time series.
However, the calculated covariance function $F_{XY}(q,s)$ (Fig. \ref{fig:WS} panel (a)) reveals a cross-over that separates two different scaling regimes, respectively related to short and long-range cross-correlations. Such a cross-over was also observed in correlated single-variable time series, as reported for instance in \cite{nicosia2013characteristic}.
Therefore, we estimate the fractal characteristics separately for each identified regime.
Differently from the above-discussed results, it is possible to estimate the fractal cross-correlations only for larger values of $q$.
This fact suggests that cross-correlations are present between large amplitudes of the signals, whereas smaller ones can be considered as fractal uncorrelated.
Additionally, $\lambda_q$ is close to $h_{xy}$, indicating strong relations between the time series in the considered range of amplitudes.
The only exception is for the short-range characteristics of the VCL-VD pair, where differences between $\lambda_q$ and $h_{xy}$ are not negligible.

By inspecting the results with more attention, it is possible to notice that the short-range cross-correlations are characterized by larger $\lambda_q$ then the long-range ones.
Weak dependence of $\lambda_q$ on $q$ suggests monofractal character of correlation. The strong relation between VCC and VD time series directly results from the procedure for generating WS graphs. In fact, random rewiring of edges lowers the average shortest path length and thus increases the vertex centrality.
The increase of vertex degree is always related with an increase of centrality, resulting in positive cross-correlations between large values of the corresponding time series.

The observed fractal cross-correlation between VCC and VCL time series is a by-product of negative correlations between centrality (degree) and clustering coefficient of the vertices; indicated with (N) in the plots shown in Fig. \ref{fig:WS} panel (a) and (b).
In fact, the rewiring procedure has a detrimental impact on the local clustering structure.
A large probability $p$ of rewiring randomizes the resulting graphs, hence lowering the (average) clustering coefficient.
For instance, for $p=0.1$ the average VCL is $\approx 0.44$, while for $p=0.5$ is $\approx 0.08$.
This is clearly visible in the evolution of the calculated characteristics. The larger $p$, the weaker the cross-correlation between VCL-VD (VCC) and the stronger relation between VCC and VD.
In the $p=0.5$ case, homogeneous scaling is observed in the covariance functions. Moreover, clustering and vertex degree are correlated only on small scales.

When $p=0.9$, instead, the cross-correlation between VCL and VD (VCC) disappears completely, since the randomization is strong and the resulting WS networks become similar to ER graphs.
Thus, the level of randomness of the networks can be directly related to the strength of fractal cross-correlation between vertex observables. This is not possible in the case of DFA algorithm, for which the small-word graphs generated with large control parameter (large degree of randomness) are not distinguishable from pure random networks.

\subsection{Protein contact networks}
\label{sec:pcn}

Proteins are large biopolymers made of amino acids arranged in sequence. After folding, such macro molecules assume a 3D conformation that can be represented as a network \cite{doi:10.1021/cr3002356}.
Vertices in such networks are the amino acid residues (although it is possible to provide also an all-atom representation). Edges are added between any two residues within a spatial distance below some suitable threshold (usually chosen in the 8--10 \AA\ range). In the literature, such networks are typically called protein contact networks (PCNs).
Here we process two sample PCNs (in the following denoted as JW0058 and JW0179) taken from \cite{mixbionets2,ecoli_graph}.

\begin{figure}[ht!]
\centering

\begin{tabular}{cc}
\includegraphics[scale=0.27,keepaspectratio=true]{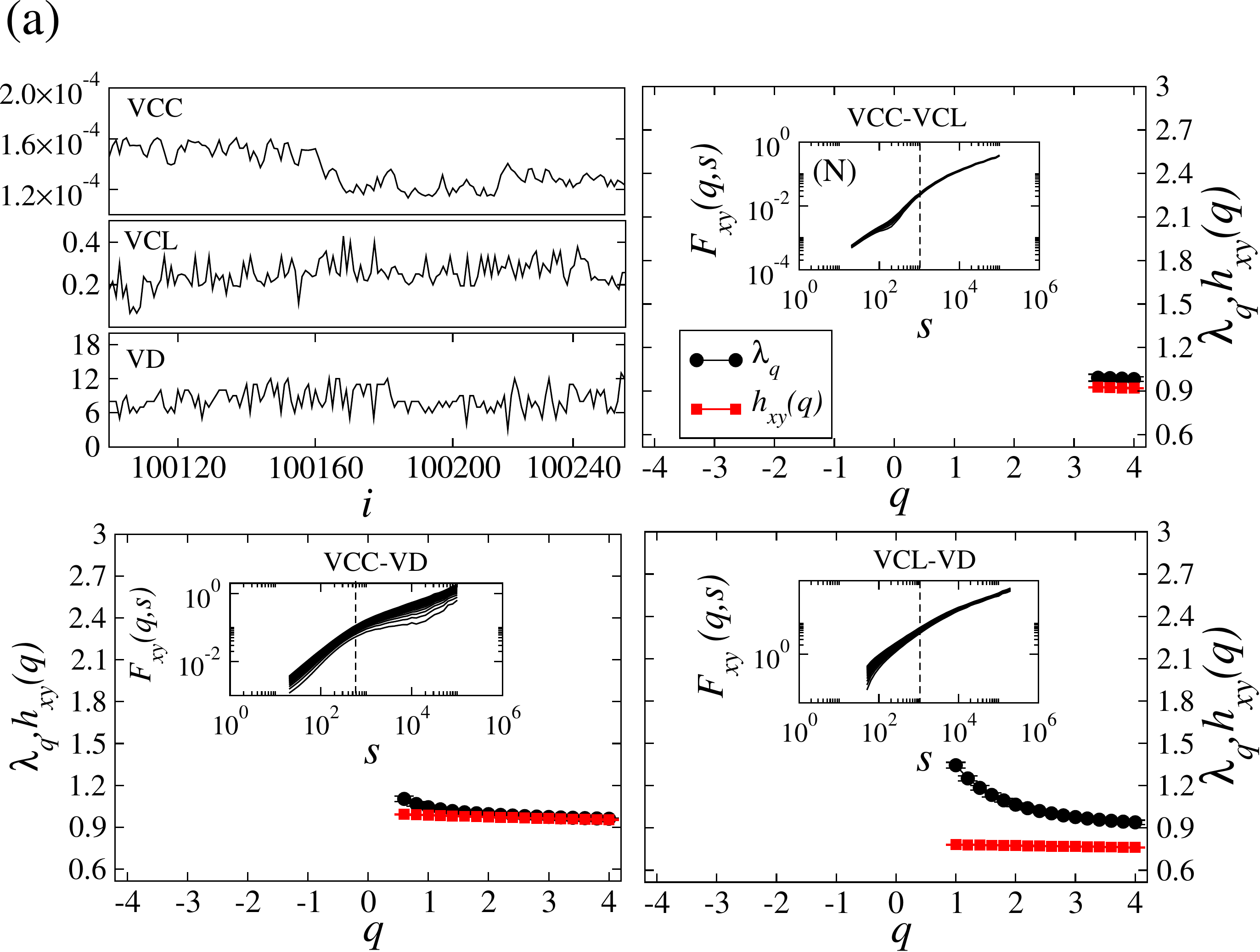}
&
\includegraphics[scale=0.27,keepaspectratio=true]{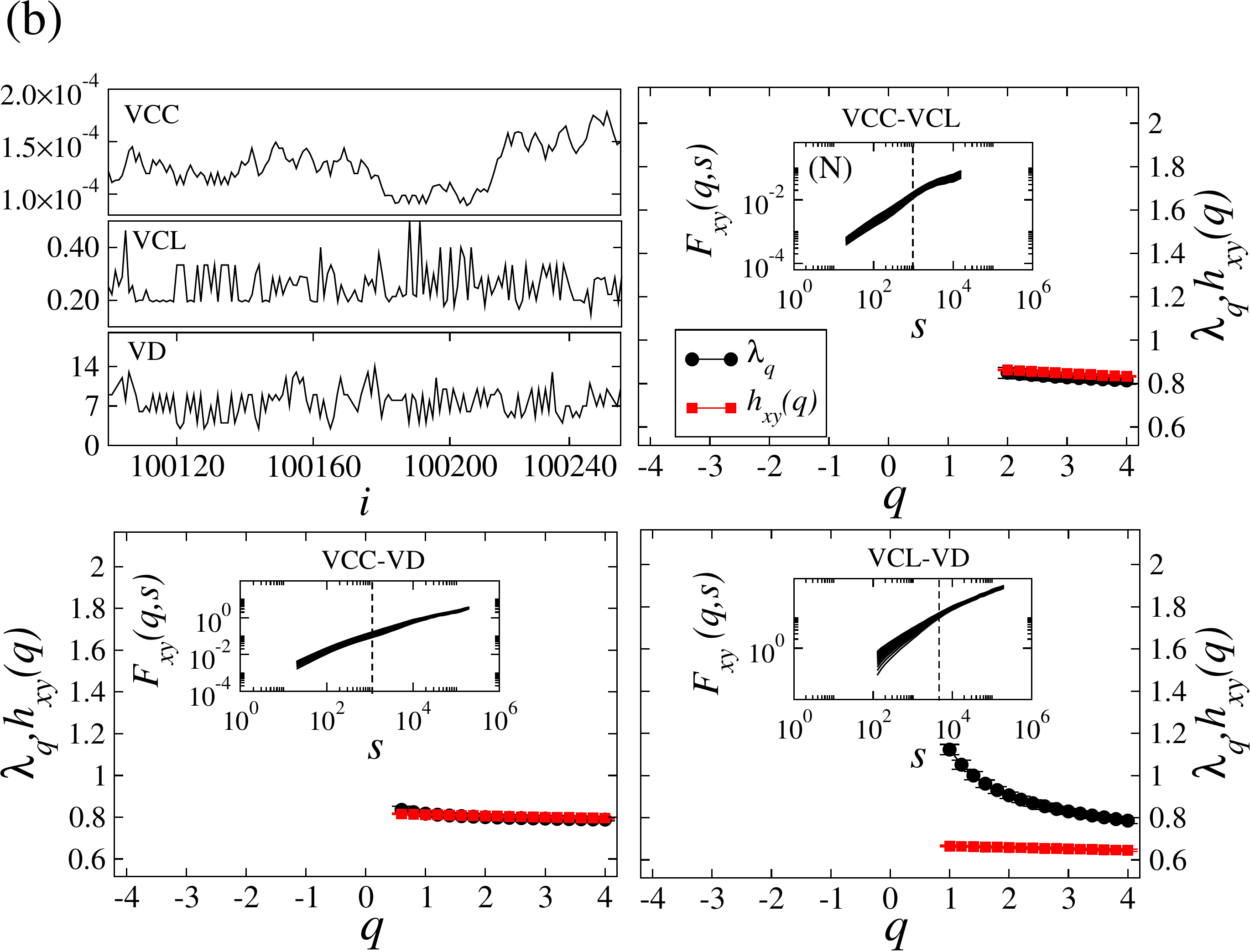}
\end{tabular}

\caption{Results for proteins JW0058 (a) and JW0179 (b).}
\label{fig:PCN}
\end{figure}

Results for scaling and $q$-dependency are shown in Fig. \ref{fig:PCN} panel (a) and (b).
The scaling properties of the fractal cross-correlation between VCC, VCL, and VD depend on the considered PCN.
In general, the scaling of the covariance function $F_{XY}(q,s)$ for JW0179 protein is more clear than the one of JW0058.
Nonetheless, it is possible to observe some common features of the temporal organization of the related two-variable time series under analysis.
The fractal character of cross-correlations is identified for all pairs of time series. From this point of view, the two PCNs denote similarity with WS graphs.
Monofractal cross-correlations are identified for VCC-VCL and VCC-VD time series.
The values of $\lambda_q$ spectrum are similar to those estimated for WS graphs with $p=0.1$ within a small-scale regime. However, the fractal cross-correlation of the VCC-VCL pair is restricted only to the largest amplitudes of the signals (largest $q$).
Differently from the above-discussed artificial networks, for PCNs we identified multifractal cross-correlations in VCL-VD pair, highlighted by the strong $q$-dependency of $\lambda_q$.
Nonetheless, the significant difference between $\lambda_q$ and $h_{xy}$ suggests weak fractal correlations between corresponding time series.

Moreover, we identified negative fractal cross-correlation for VCC-VCL and positive one for VCL-VD time series for both PCNs. The negative cross-correlation between closeness centrality and clustering coefficient, as shown above in Fig. \ref{fig:WS} panel (a) and (b), is characteristic for WS graphs and originates from long-range contacts existing in the network. In turn, positive fractal cross-correlation between clustering and vertex degree characterizes BA graphs, since they contain hubs.
This result indicates that the structure of PCNs possesses a dual nature and carries characteristics of both BA and WS networks.

In the PCNs case, the negative cross-correlations for VCC-VCL can be further justified as follows.
Proteins are physical objects made of amino acids that differ in terms of volume (and mass).
This observation implies the existence of specific structural (physical) constraints in PCNs.
The average clustering coefficient (closeness centrality) for JW0058 is 0.265 (0.112) and 0.277 (0.115) for JW0179.
On the other hand, for WS graphs generated with $p=0.05$ and 0.1 we have 0.525 (0.127) and 0.447 (0.162), respectively.
The significantly lower average clustering coefficient (with comparable values of average closeness centrality) is a direct consequence of the aforementioned packing constraints existing in proteins. In fact, clustering coefficient cannot grow arbitrarily due to the limitation imposed on the number of possible neighbours.
Therefore, we suggest that, in the PCNs case, the negative cross-correlation for VCC-VCL arises also as a consequence of the packing constraints influencing the structure of PCNs.

The length of autocorrelation in all the time series considered in this study ranges between about 5 time steps for ER and BA networks up to 300-400 for WS and PCNs. The used time series are thus orders of magnitude longer than the range of autocorrelation, so the statistical sampling is equally robust in all the cases taken into account.

In order to fully verify the significance of the results, we repeated our analysis by randomly shuffling all original time series (results not shown). Shuffling destroys all temporal correlations existing in time series but preserves statistical distribution of the data. In all cases, covariance function strongly fluctuates around zero, suggesting the lack of fractal cross-correlation and hence supporting the reliability of our findings.

\section{Conclusions}
\label{sec:conclusions}

In this paper, we investigated the scaling properties of cross-correlations calculated for two-variable time series containing vertex properties elaborated from complex networks.
We have considered both models and real complex networks (protein contact networks).
The developed methodological approach consists in mapping a network to a two-variable time series by performing random walks on the vertices. At each time step, two distinct vertex properties are emitted, forming thus two-variable time series encoding topological information regarding the network under analysis.

Our study showed that the MF-CCA methodology is able to distinguish between the most common types of artificially generated graphs.
In this respect, the most complex structure of the cross-correlation is related with Watts-Strogatz networks (with small probability of rewiring $p$).
On the other hand, random networks represented in our analyses by Erd\"{o}s-R\'{e}nyi and Watts-Strogatz (with large $p$) graphs are characterized only by the monofractal cross-correlation between vertex degree and centrality. Complexity of Barab\'{a}si-Albert networks is located somewhere between Erd\"{o}s-R\'{e}nyi and Watts-Strogatz graphs.
However, the most clear example of the strength of the MF-CCA methodology is related to the analysis of the protein contact networks.
In this case, obtained cross-correlation characteristics suggest the existence of a structural duality for such networks.
In particular, protein contact networks can be considered as a compromise of Barab\'{a}si-Albert and Watts-Strogatz graphs (with small $p$), although with a much more ``democratic'' distribution of links than in pure Barab\'{a}si-Albert and higher assortativity with respect to Watts-Strogatz models.

Our findings suggest that the universal methodology based on MF-CCA is useful to investigate the structure and function of complex networks in terms of time series analysis, by quantifying effects that are undetectable by means of other previously used methods.

\bibliographystyle{apsrev4-1}
\bibliography{Bibliography.bib}
\end{document}